\documentclass[%
reprint,
 amsmath,amssymb,
 aps,
]{revtex4-1}

\usepackage{color}
\usepackage{graphicx}
\usepackage{dcolumn}
\usepackage{bm}

\def\be{\begin{equation}}
\def\ee{\end{equation}}

\begin{document}

\preprint{APS/123-QED}

\title{A self-consistent study of magnetic field effects on hybrid stars}

\author{B. Franzon}
\affiliation{Frankfurt Institute for Advanced Studies,
Ruth-Moufang - 1 60438, Frankfurt am Main,
Germany}
\email{franzon@fias.uni-frankfurt.de}

\author{V. Dexheimer}
\affiliation{Department of Physics, Kent State University, Kent OH 44242 USA}

\author{S. Schramm}
\affiliation{Frankfurt Institute for Advanced Studies,
Ruth-Moufang -1 60438, Frankfurt am Main,
Germany}

\date{\today}

\begin{abstract}
In this work we study the effects of strong magnetic fields on hybrid stars by using a full  general-relativity approach,  solving the coupled Maxwell-Einstein equation in a self-consistent way. The magnetic field is assumed to be axi-symmetric and poloidal. We take into consideration the anisotropy of the energy-momentum tensor due to the magnetic field, magnetic field effects on equation of state, the interaction between matter and the magnetic field (magnetization), and the anomalous magnetic moment of the  hadrons. The equation of state used is an extended hadronic and quark SU(3) non-linear realization of the sigma model that describes magnetized hybrid stars containing nucleons, hyperons and quarks. According to our results, the effects of the magnetization and  the magnetic field on the EoS do not play an important role on global properties of these stars.  On the other hand,  the magnetic field causes the central density in these objects to be reduced, inducing major changes in the populated degrees of freedom and, potentially, converting a hybrid star into a hadronic star.  
\end{abstract}

\pacs{Valid PACS appear here}
\maketitle


\section{Introduction}

Neutron stars  undoubtedly belong to the most suitable environments for studying properties of strongly interacting matter under extreme conditions. For example, the density inside these objects can reach values much higher than the nuclear saturation density $\sim 2.7 \times 10^{14}$ $\mathrm{g/cm^{3}}$. This makes neutron stars natural laboratories where one can examine and shed some light on the still open question concerning the equation of state (EoS) for ultra-dense matter and the role played by exotic degrees of freedom, such as hyperons and quarks. 
  
Another important feature present in compact objects and studied in this work is their strong magnetic field.  From pulsar observations, the magnitude of the  surface magnetic field in neutron stars has been found to be generally of the order of $\mathrm{10^{12} - 10^{13}\, G}$.  However, according to observations of star periods and period derivatives, classes of neutron stars known  as Anomalous X-ray Pulsars and Soft-Gamma-Ray-Repeaters can have surface magnetic field as large as $\mathrm{10^{14} -10^{15}\, G }$. These are usually referred to as magnetars \cite{Duncan:1992hi,Thompson:1993hn, Thompson:1996pe,  paczynski1992gb, melatos1999bumpy}
. One expects to find even stronger magnetic fields inside these stars as already calculated in Ref.~\citep{Makishima:2014dua}. According to virial theorem arguments, which give an upper estimate for the magnetic inside neutron stars, they can possess central magnetic fields as large as $10^{18-20}$ G \cite{ferrer2010equation, lai1991cold, fushiki1989surface, cardall2001effects}. 

The origin of strong magnetic fields in compact stars  is still unclear. One common hypothesis involves the flux conservation of the progenitor magnetic field \cite{woltjer}. However, this idea is not suitable for magnetars since a canonical neutron star $\rm{M \sim 1.4\, M_{\odot}}$ would require a radius less than its Schwarzschild radius in order to generate a surface magnetic field of the order of $\mathrm{10^{15}G}$ \citep{tatsumi2000ferromagnetism}.  Another possibility suggested by Thompson and Duncan in Ref. \citep{Duncan:1992hi} describes a newly born neutron star combining convection and differential rotation to generate a dynamo process which is able to generate fields as large as $\mathrm{10^{15}\, G }$. However, this standard explanation fails when trying to explain the supernova remnants associated with these objects \citep{vink2006supernova}. 

Whatever the origin of strong magnetic fields might be, they affect locally the microphysics  of the equation of state (EoS), as for example, due to the Landau quantization of the energy levels of charged particles and the effect of the anomalous magnetic moment (AMM) of charged and uncharged particles. Globally, magnetic fields affect the structure of neutron stars  through the Lorentz force associated with the macroscopic currents that generate the field. They also affect the structure of the spacetime, as magnetic fields are now a source for the gravitational field through the Maxwell energy-momentum tensor. As a consequence, magnetized stars are anisotropic and require a general-relativity treatment beyond the solution of the Tolman-Oppenheimer-Volkoff (TOV) equations \cite{tolman1939static, oppenheimer1939massive}. 

In this work, we model magnetized hybrid stars in a self-consistent way  by solving Einstein-Maxwell equations in the same way as done in Refs.~\cite{Bonazzola:1993zz, Chatterjee:2014qsa}. In Ref.~\cite{Chatterjee:2014qsa}, the authors studied magnetized quark stars taking into account the magnetic field in the equation of state, the magnetization term for the matter, and also the magnetic field in the gravitational field equations.  They have found that neither the magnetic field nor the magnetization  change significantly the global properties of these stars for a magnetic field strength of the order of $\sim$ $10^{18}\,$G. Note that, the equation of state  used by those authors, namely, a quark CFL model \cite{noronha2007color} possesses a high baryon density range reaching  two times the saturation density  at the surface  of star.     

 In order to assess the  role that a magnetic field dependent equation of state and the magnetization play in the global properties of stars, we use in this work a more complex equation of state than in Ref.~\cite{Chatterjee:2014qsa}. Our EoS describes magnetized hybrid stars containing nucleons, hyperons and quarks and  takes into account the anomalous magnetic moment for all hadrons.  As a consequence, it produces a magnetization much higher than the one used in the Ref.~\cite{Chatterjee:2014qsa}. Despite this, we show that the neutron star structure, like its mass-radius relationship, is not modified drastically with the inclusion of the magnetic field in the EoS and the magnetization.  On the other hand, the particle population is significantly modified when the magnetic field is included. The main impact observed is the conversion of a non-magnetized hybrid star with hadron and quark degrees of freedom to a highly magnetized hadronic star composed simply by nucleons. In reality, the temporal star evolution goes in the other direction, as the magnetic field of the star decays over time allowing a hadronic star to become a hybrid one. 

\section{Formalism}
 
In this work, we study stationary highly magnetized neutron stars. In this context, stationary neutron stars with no magnetic-field-dependent EoS were studied in Refs.~\cite{Bonazzola:1993zz, Bocquet:1995je}.   An extension of these works,  where the authors included magnetic fields effects in the EoS, was presented in Ref.$\,$\cite{Chatterjee:2014qsa}. Here, we follow the same general relativity formalism  and setup as in Ref.$\,$\cite{Chatterjee:2014qsa}.
Details of the relevant equations, numerical procedure and tests  can be found in the references cited above. We present here only the key equations that are solved numerically for the sake of completeness and better understanding for the reader.
 
 The choice of the coordinates in general relativity is crucial not only to write the gravitational equations in an advantageous form but also to make them  easier to solve numerically. In the present case,  due to the symmetry of the system,  a polar-spherical type coordinate system is chosen, namely,  the Maximal-Slicing-Quasi-Isotropic coordinates (MSQI) (for a review see Ref.~\cite{gourgoulhon20123+}).  The metric in the MSQI coordinate system is written as:
\begin{align}
ds^2 =  &-N^{2}dt^{2} + B^{2}r^{2}\sin^{2}\theta(d\phi - N^{\phi} dt)^{2}   \nonumber\\
&- A^2(dr^2 + r^2 d\theta^2),
\label{metric}
\end{align}
with $N^{\phi} (r, \theta)$ being the shift vector, $N (r, \theta)$ the lapse function and A and B  functions of $r$ and $\theta$.   
The energy-momentum tensor of the system  reads:
\begin{align}
T^{\mu\nu} = & (e+p)u^{\mu}u^{\nu} + pg^{\mu\nu}  \nonumber\\
 & + \frac{m}{B} \left (  b^{\mu}b^{\nu}   -  (b\cdot b) (u^ {\mu} u^{\nu} +  g^{\mu\nu})  \right)\nonumber\\
 &+ \frac{1}{\mu_{0}} \left (  -b^{\mu}b^{\nu}   + (b\cdot b) u^ {\mu} u^{\nu} +  \frac{1}{2} g^{\mu\nu} (b\cdot b)   \right) ,
 \label{emt}
\end{align}
where $m$ and $B$ are the lengths of the magnetization and magnetic field 4-vectors. 
Here, the first and second terms correspond to the matter contribution. 
They were written separating the isotropic matter contribution (first term) from the magnetization contribution which is anisotropic (second term). The third term in Eq.~\eqref{emt} is the pure electromagnetic contribution to the energy-momentum tensor. 
The energy density is $e$, the isotropic contribution to the pressure is $p$ and the fluid velocity is $u^{\mu}$. 
The magnetic field in the fluid rest frame is $b^{\mu} = \rm{diag}\, (0,0,0,B)$, where $B$ is defined to point into the z-direction. 
The metric follows the convention $g^{\mu\nu} = \rm{diag}\, (-1,1,1,1)$. In the rest frame of the fluid (using $u^{\mu} = \rm{diag}\, (1,0,0,0)$) we can write: 

\begin{align}
 & T^{0}_{\!m\,0} = e,  \nonumber\\
 & T^{1}_{\!m\,1} = p - mB, \nonumber\\
 & T^{2}_{\!m\,2} = p - mB, \nonumber\\
 & T^{3}_{\!m\,3} = p,
\label{mattercontribution}
\end{align}
where the components  $T^{1}_{\!m\,1}$ and  $T^{2}_{\!m\,2}$ are usually referred to as perpendicular or transversal pressure (with respect to $B$) and $T^{3}_{\!m\,3}$ as parallel or longitudinal pressure, respectively.  The third term in Eq.~\eqref{emt} corresponds to the pure magnetic field contribution. It can be written as:
\begin{align}
 & T^{0}_{\!\!B\,0} = \frac{B^2}{2\mu_{0}},  \nonumber\\
 & T^{1}_{\!\!B\,1} = \frac{B^2}{2\mu_{0}}, \nonumber\\
 & T^{2}_{\!\!B\,2} = \frac{B^2}{2\mu_{0}}, \nonumber\\
 & T^{3}_{\!\!B\,3} = -\frac{B^2}{2\mu_{0}}.
\label{fieldcontribution}
\end{align}
Following the notation of Ref.~\cite{cardall2001effects}, the standard stress-energy tensor for the perfect fluid (PF) gives an energy density:
\be
E^{(PF)}= \Gamma^2 (e + p) - p,
\label{energydensitypf}
\ee
while the momentum density flux is given by:
\be 
J^{(PF)}_{\phi} = \Gamma^2 (e + p) U .
\label{momentumdensitypf}
\ee
The  3-tensor stress components can be written as:
\be
S^{(PF) r}_{\;\;\; r} = S^{(PF) \theta}_{\;\;\; \theta}  = p ,
\ee
and 
\be
S^{(PF)\phi}_{\;\;\; \phi} = p + (E^{(PF)} + p) U^2,
 \ee
where the Lorentz factor is given by $\Gamma = (1 - U^2)^{-\frac{1}{2}}$ being $U$ the fluid velocity  defined  as:
\be
U = \frac{Br\sin\theta}{N}(\Omega - N^\phi),
\ee
with the lapse function $N^\phi$ and  the angular velocity $\Omega$, as measured by an observer at infinity (see Ref.~\cite{cardall2001effects} for more details). 
The expressions for electric and magnetic field components as measured by the observer ($\mathcal{O}_{0}$) with $n^{\mu}$ velocity can be written as \cite{lichnerowicz1967relativistic}:
\begin{align}
E_{\alpha}&= \nonumber F_{\alpha\beta}n^{\beta} \\ 
\hspace*{-2cm}&=\left(0,\frac{1}{N}\left[\frac{\partial A_{t}}{\partial r}+N^{\phi}\frac{\partial A_{\phi}}{\partial r}\right ],\frac{1}{N}\left[\frac{\partial A_{t}}{\partial \theta}+N^{\phi}\frac{\partial A_{\phi}}{\partial \theta}\right],0\right),
\label{electricfield}
\end{align}
\begin{align}
\hspace{-0.8cm}B_{\alpha}&=\nonumber -\frac{1}{2} \epsilon_{\alpha\beta\gamma\sigma} F^{\gamma\sigma}n^{\beta} \\
\hspace{-0.8cm}&= \left( 0 , \frac{1}{ B r^{2} \sin \theta} \frac{\partial A_{\phi}}{\partial \theta}, - \frac{1}{ B \sin \theta} \frac{\partial A_{\phi}}{\partial r} , 0  \right).
\label{magneticfield}
\end{align}
The corresponding energy density (electromagnetic $+$ magnetization) reads:
\be 
E^{(EM)} = \frac{1}{2\mu_{0}} \left[ \left(1 + 2 \frac{m}{B}\right)E^i E_{i}  + B^i B_{i}      \right],
\label{energydensity}
\ee
and the momentum density flux can be written as:
\be 
J^{(EM)}_{\phi} = \frac{1}{\mu_{0}} \left[ A^2(B^r E^\theta  - E^r B^\theta) + \frac{m}{B} B^i B_{i} U     \right].
\label{momentumdensity}
\ee
The 3-tensor stress components are given by:

\begin{align}
S^{(EM) r}_{\;\;\; r} = \nonumber & \frac{1}{2\mu_{0}} (E^\theta E_{\theta}  -    E^r E_{r} +  B^\theta B_{\theta} -  B^r B_{r}) + \\
 &+ \frac{2m}{B}  \frac{B^\theta B_{\theta}}{\Gamma^2},
\label{stress1}
\end{align}
\begin{align}
S^{(EM) \theta}_{\;\;\; \theta} = \nonumber & \frac{1}{2\mu_{0}} (E^r E_{r}  -   E^\theta E_{\theta} +  B^r B_{r}  -  B^\theta B_{\theta}) + \\
& + \frac{2m}{B}  \frac{B^{r}B_{r}}{\Gamma^2},
\label{stress2}
\end{align}
\begin{align}
\hspace{0.7cm}S^{(EM)\phi}_{\;\;\; \phi} = \frac{1}{2\mu_{0}} \left[ E^i E_{i}+ B^i B_{i} + \frac{2m}{B} (1 + \Gamma^2 U^2) \frac{B^iB_{i}}{\Gamma^2} \right].
\label{stress3}
\end{align}

Besides the above definitions,  other important quantities measured by Eulerian obeserver are the $\it{circunferencial \, radius }\,\, \rm{R_{circ}}$:
\be 
R_{circ} =   B ( r_{eq},\frac{\pi}{2} ) r_{eq},
\label{rcir}
\ee
being $r_{eq}$ the coordinate equatorial radius. The total gravitational mass of the star $M$ is given by:
\begin{align}
&M =  \int A^{2}Br^{2} \times \nonumber\\
    &\left[ N(E+S)+ 2N^{\phi}B(E+p) Ur\sin\theta \right] \sin\theta dr d\theta d\phi,
\label{mass}
\end{align}
with $E$ being the total energy density of the system $E = E^{(PF)} +   E^{(EM)}$ and  $S^{i}_{i} $ total stress $S^{i}_{i}  = S^{(PF) \, i}_{i} + S^{(EM) \, i}_{i}   $, where i = $r$, $\theta$ and $\phi$.




Assuming that the matter inside the star has infinite conductivity, the electric field measured by the coming observer $\mathcal{O}_{1}$ (observer with velocity $u^{\mu}$) must be zero, i.e. 	$E^{\prime}_{\alpha} = u^{\beta}F_{\alpha\beta} = 0$.  This condition leads to the following relation of the  4-vector potential inside the star:
\be 
\frac{\partial A_{t}}{\partial x^{i}} = - \Omega \frac{\partial A_{\phi}}{\partial x^{i}},
\ee
which for the case of rigid rotation, $\Omega = const$, one has:
\be 
A_{t} = - \Omega A_{\phi}.
\label{potentialrelation}
\ee
As calculated in Ref.~\cite{Bonazzola:1993zz}, the equation of motion ($\nabla_{\mu}T^{\mu\nu}= 0$) reads
\be 
\frac{\partial}{\partial x^{i}} \left( H + \nu - ln \Gamma \right) - \frac{1}{e+p} \left( j^{\phi} - \Omega j^{t} \right) \frac{\partial A_{\phi}}{\partial x^{i}} = 0,
\label{efm}
\ee
with $\nu = \nu(r,\theta):=ln (N)$. The relativistic log-enthalpy $H(r,\theta)$ in Eq.~\eqref{efm} is written as:
\be
H:= ln \left( \frac{e+p}{m_{b}n_{b}c^{2}  } \right),
\label{enthalpy}
\ee
where $m_{b}$ denotes the mean baryon mass $1.66\times10^{-27} \,\rm{kg}$ and $n_{b}$ the baryon number density, respectively.
As shown in Ref.~\cite{Bonazzola:1993zz}, one can relate the electric current to the electromagnetic potential through the current function\, $f$:
\be 
j^{\phi} - \Omega j^{t} = \left( e + p \right) f \left( A_{\phi} \right),
\label{totalcurrent}
\ee
and by integration of the equation of motion ($\nabla_{\mu}T^{\mu\nu}= 0$), one obtains:
\be 
H \left(r, \theta \right) + \nu \left(r, \theta \right) -ln \Gamma \left( r, \theta \right) + M \left(r, \theta \right) = const,
\label{equationofmotion}
\ee
being  $M(r,\theta)$ the electromagnetic term induced by the Lorentz force \cite{Bocquet:1995je} written as:
\be 
M \left(r, \theta \right) = M \left( A_{\phi} \left(r, \theta \right) \right): = - \int^{0}_{A_{\phi}\left(r, \theta \right)} f\left(x\right) \mathrm{d}x.
\ee
The model built in this work used a current function $f=f_{0}=const$. As shown in Ref.~\cite{Bocquet:1995je}, other choices for $f$ are possible, however, they do not alter the conclusions qualitatively. For different values of $f_{0}$, the electric current changes (see Eq.~\eqref{totalcurrent}) and, therefore, the intensity of the magnetic field in the star changes. 

\section{Magnetized Equation of State}

When, in addition to a usual hadronic phase, neutron star models also include a quark phase, they are normally described by two different equations of state. These equations of state are connected at the point in which the pressure of the quark phase becomes higher than the pressure of the hadronic phase. 
In our approach, we have instead a combined model of hadrons and quarks and one equation of state for both phases. In this case, we can study important features of the deconfinement phase transition,  like the strength of the transition, the mixing of phases and also the accompanying chiral symmetry restoration.
For this purpose, we extended the hadronic SU(3) non-linear realization of the sigma model \cite{PhysRevC.88.014906,Papazoglou:1998vr,Dexheimer:2008ax,Dexheimer:2009hi} to include quark degrees of freedom in a spirit similar to the PNJL model  \cite{Fukushima:2003fw}, in the sense that it uses the Polyakov loop $\Phi$ as the order parameter for deconfinement. The Lagrangian density of the model in mean field approximation reads:
\begin{eqnarray}
&\mathcal{L} = \mathcal{L}_{mag}+\mathcal{L}_{Kin}+\mathcal{L}_{Int}+\mathcal{L}_{Self}+\mathcal{L}_{SB}-U,&
\end{eqnarray}
where besides the kinetic energy term for hadrons, quarks, and leptons (included to insure charge neutrality),
the terms:
\begin{eqnarray}
&\mathcal{L}_{mag}=-\sum_i \bar{\psi_i} (q_i e \gamma^\mu A_\mu + \frac 1 2 \kappa \sigma^{\mu\nu} F_{\mu\nu}) \psi_i,\nonumber&\\&
\label{5}
\end{eqnarray}
\begin{eqnarray}
&\mathcal{L}_{Int}=-\sum_i \bar{\psi_i}[\gamma_0(g_{i\omega}\omega+g_{i\phi}\phi+g_{i\rho}\tau_3\rho)+M_i^*]\psi_i,\nonumber&\\&
\end{eqnarray}
\begin{eqnarray}
&\mathcal{L}_{Self}=\frac{1}{2}(m_\omega^2\omega^2+m_\rho^2\rho^2+m_\phi^2\phi^2)\nonumber&\\&
+g_4\left(\omega^4+\frac{\phi^4}{4}+3\omega^2\phi^2+\frac{4\omega^3\phi}{\sqrt{2}}+\frac{2\omega\phi^3}{\sqrt{2}}\right)\nonumber&\\&-k_0(\sigma^2+\zeta^2+\delta^2)-k_1(\sigma^2+\zeta^2+\delta^2)^2&\nonumber\\&-k_2\left(\frac{\sigma^4}{2}+\frac{\delta^4}{2}
+3\sigma^2\delta^2+\zeta^4\right)
-k_3(\sigma^2-\delta^2)\zeta&\nonumber\\&-k_4\ \ \ln{\frac{(\sigma^2-\delta^2)\zeta}{\sigma_0^2\zeta_0}},&
\end{eqnarray}
\begin{eqnarray}
&\mathcal{L}_{SB}= -m_\pi^2 f_\pi\sigma-\left(\sqrt{2}m_k^ 2f_k-\frac{1}{\sqrt{2}}m_\pi^ 2 f_\pi\right)\zeta,\nonumber&\\&
\end{eqnarray}
\begin{eqnarray}
&U=(a_0T^4+a_1\mu^4+a_2T^2\mu^2)\Phi^2&\nonumber\\&+a_3T_0^4\log{(1-6\Phi^2+8\Phi^3-3\Phi^4)},&
\end{eqnarray}
represent the magnetic and anomalous magnetic moment (AMM) interactions with the fermions, the interactions between baryons or quarks and vector and scalar mesons, the self interactions of scalar and vector mesons, an explicit chiral symmetry breaking term (responsible for producing the masses of the pseudo-scalar mesons), and a potential $U$ for the $\Phi$ field. The later is important in order to reproduce a realistic structure for the QCD phase diagram over the whole range of chemical potentials and temperatures, including realistic thermodynamic behaviour at vanishing chemical potential as shown in Ref.~\cite{Dexheimer:2008ax}.

In Eq.~\eqref{5}, $q_i$ is the electric charge of each particle in multiples of the electron charge $e$, $A_\mu$ is the electromagnetic field, $\kappa$ represents the tensorial coupling strength of baryons with the electromagnetic field tensor, and $\sigma^{\mu\nu}=i[\gamma^\mu,\gamma^\nu]/2$ . The mesons included are the vector-isoscalars $\omega$ and $\phi$ (vector meson with hidden strangeness), the vector-isovector $\rho$, the scalar-isoscalars $\sigma$ and $\zeta$ (scalar meson with hidden strangeness) and  the scalar-isovector $\delta$, with $\tau_3$ being twice the isospin projection of each particle. The isovector mesons affect isospin-asymmetric matter and are, consequently, important for neutron star physics. The coupling constants of the model can be found in Ref.~\cite{Dexheimer:2008ax,Dexheimer:2009hi}. The hadronic sector was fitted to reproduce the vacuum masses of the baryons and mesons, nuclear saturation properties, reasonable values for the hyperon potentials and the pion and kaon decay constants ($f_{\pi}$ and $f_{k}$). The quark sector was fitted to reproduce lattice QCD data at vanishing chemical potential and phase diagram information, such as the location of the critical end-point and a continuous first-order phase transition line that terminates on the zero temperature axis at around  four times saturation density.

The effective masses of the baryons and quarks are generated by the scalar mesons except for a small explicit
mass term $M_0$ and the term containing $\Phi$:
\begin{eqnarray}
&M_{B}^*=g_{B\sigma}\sigma+g_{B\delta}\tau_3\delta+g_{B\zeta}\zeta+M_{0_B}+g_{B\Phi} \Phi^2,&
\label{6}
\end{eqnarray}
\begin{eqnarray}
&M_{q}^*=g_{q\sigma}\sigma+g_{q\delta}\tau_3\delta+g_{q\zeta}\zeta+M_{0_q}+g_{q\Phi}(1-\Phi).\nonumber&\\&
\label{7}
\end{eqnarray}
With the increase of temperature and/or density, the $\sigma$ field (non-strange chiral condensate) decreases in value, causing the effective masses of the particles to decrease towards chiral symmetry restoration. The field $\Phi$ assumes non-zero values with the increase of temperature/density and, due to its presence in the baryons effective mass (Eq.~\eqref{6}), suppresses their presence. On the other hand, the presence of the $\Phi$ field in the effective mass of the quarks, included with a negative sign (Eq.~\eqref{7}), ensures that they will not be present at low temperatures/densities. In this way, the interaction with the medium determines which are the degrees of freedom present in the system.

The magnetic field in the z-direction forces the energy eigenstates in the x and y directions of the charged particles to be quantized into Landau levels $\nu$:
\begin{equation}
E_{i_{\nu s}}^*=\sqrt{k_{z_{i}}^2+\left(\sqrt{M_i^{*2}+2\nu|q_i|B}-s_i\kappa_i B \right)^2},
\end{equation}
where $k_i$ is the fermi momentum and $s_i$ the spin of each fermion. The last term comes from the anomalous magnetic moment (AMM) of the particle that splits the energy levels with respect to the alignment/anti-alignment of the spin with the magnetic field. The AMM also modifies the energy levels of the uncharged particles
\begin{equation}
E_{i_{s}}^*=\sqrt{k_{i}^2+\left({{M_i^*}^2}-s_i\kappa_i B^* \right)^2}.
\end{equation}
The AMM constants $\kappa_i$ have values $\kappa_p=1.79$, $\kappa_n=-1.91$, $\kappa_\Lambda=-0.61$, $\kappa_\Sigma^+=1.67$, $\kappa_\Sigma^0=1.61$, $\kappa_\Sigma^-=-0.38$, $\kappa_\Xi^0=-1.25$, $\kappa_\Xi^-=0.06$. The sign of $\kappa_i$  determines the preferred orientation of the spin with the magnetic field. For zero temperature, the sum over the Landau levels $\nu$ runs up to a maximum value, beyond which the momentum of the particles in the z-direction would be imaginary
\begin{equation}
\nu_{max} =\frac{{E_{i_s}^*}^2+s_i\kappa_i B - {M_i^*}^2 }{2|q_i|B}.
\end{equation}

We choose to include in our calculations the AMM effect for the hadrons only, since the coupling strength of the particles $\kappa_i$ depends on the corresponding magnetic moment, that up to now is not fully understood for the quarks. Furthermore, it is stated in Ref.~\cite{Weinberg:1990xm}, that quarks in the constituent quark model have no anomalous magnetic moment, and in Ref.~\cite{PhysRevD.91.085041}, that the AMM of quarks from one-loop fermion self-energy is very small. For calculations including AMM effects for the quarks see Refs.~\cite{Chakrabarty:1996te,Suh:2001tr,PerezMartinez:2005av,Felipe:2007vb}. The AMM for the electrons is also not taken into account as its effect is negligibly small. Properties of the magnetized SU(3) non-linear realization of the sigma model were presented in Refs.~\cite{Dexheimer:2011pz,Dexheimer:2012qk} for an effective (ad hoc) variation of the magnetic field inside the star.

For this work, the equation of state for charge neutral chemically-equilibrated matter was calculated at zero temperature and over a wide interval of densities and magnetic fields with very small steps. This 2-dimensional table was included in the Lorene 
C++ class library for numerical relativity \cite{Bonazzola:1993zz, Bocquet:1995je} in order for the code to find the correct magnetic field for each density in each direction of the star and, then, find the corresponding values for the system's thermodynamical properties. Our tabulated equation of state is available upon request.

\section{ Results}

The equilibrium configurations are determined by the central enthalpy $H_{c}$ and the choice of the magnetic dipole moment $\mu$.   We could have, instead, chosen a fixed current function $f_{0}$ and allow the magnetic dipole moment to vary. We chose the former in order to have a better control of the parameter space and to investigate exclusively the effect of the magnetic field on the star. In order to do so, we built equilibrium sequences for  fixed magnetic dipole moments $\mu$ defined as (see Ref.~\cite{Bonazzola:1993zz}):
\be 
\frac{2\mu cos\theta}{r^{3}} = B_{(r)}\mid_{r\rightarrow \infty},
\label{mm}
\ee
which is simply the radial component (the orthonormal one) of the magnetic field of a magnetic dipole seen by an observer at infinity.

\begin {figure}[!h]
\begin{center}
\includegraphics[width=0.7\textwidth,angle=-90,scale=0.5]{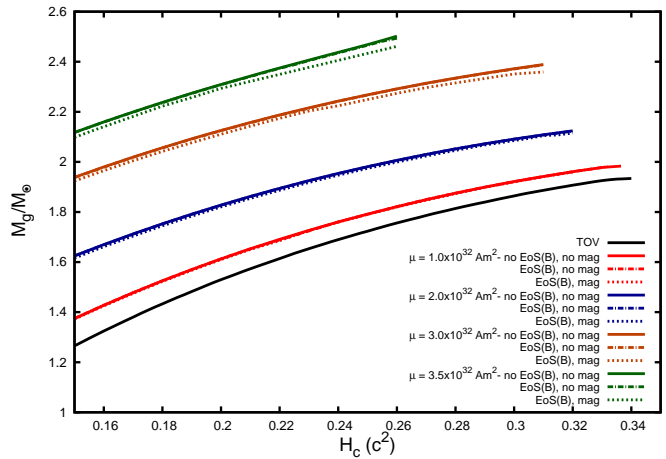}
\caption{\label{mr_nocrust_b_magmom} Relation between the gravitational mass and the central enthalpy for non-magnetized and magnetized models. In the last case, we also include the effects of the magnetic field into the equation of state (EoS(B)) and the magnetization term (mag) .}
\end{center}
\end {figure}

The magnetization is defined as $M = -\partial \Omega /\partial B$ (for more details see Ref.~\cite{Strickland:2012vu}) and in Figure~\ref{mr_nocrust_b_magmom} we show:
i) the case when the magnetic field is included only in the structure of the star (no EoS(B), no mag); ii) the effect of the magnetic field also into the equation of state on the neutron star structure without the magnetization term (EoS(B), no mag) and iii) the effect of the magnetic field also into the equation of state on the neutron star structure plus the magnetization term (EoS(B), mag). We also show the non-magnetized cased denominated TOV.

The results present in Ref.~\cite{Chatterjee:2014qsa} show that the effects of the magnetic field in the equation of state and the magnetization on the neutron star structure are not considerable. Note, however, that those authors presented solutions for magnetized quark stars whose equation of state (CFL model) is very stiff and reaches, for instance,  a number baryon density of about 2 times the nuclear saturation density at the surface of the star. In this work, we model not only the quark phase with up, down quarks, but also the hadronic phase containing the whole baryon octet, in a self-consistent way in the presence of the magnetic field.


The results in Figure~\ref{mr_nocrust_b_magmom} corroborate in part the results presented in Ref.~\cite{Chatterjee:2014qsa} showing also that the magnetic field does not have a considerable effect on the maximum mass and radius of highly magnetized neutron stars through the effect on the equation of state for given magnetic fields.  Only a very small reduction of the star mass is seen for the largest magnetic moment used in the most massive star calculated.  We see, however, from our calculation a difference in the curves when the magnetization term is included. This is due to the fact that, in our case, the magnetization strength can reach a value of about 10 times the value in Ref.~\cite{Chatterjee:2014qsa}. Note that the effect of the magnetization is to decrease stellar masses. This fact is related to the negative sign in Eq.~\eqref{mattercontribution}.

\begin {figure}[!t]
\begin{center}
\hspace*{-1cm}\includegraphics[width=0.8\textwidth,angle=-90,scale=0.5]{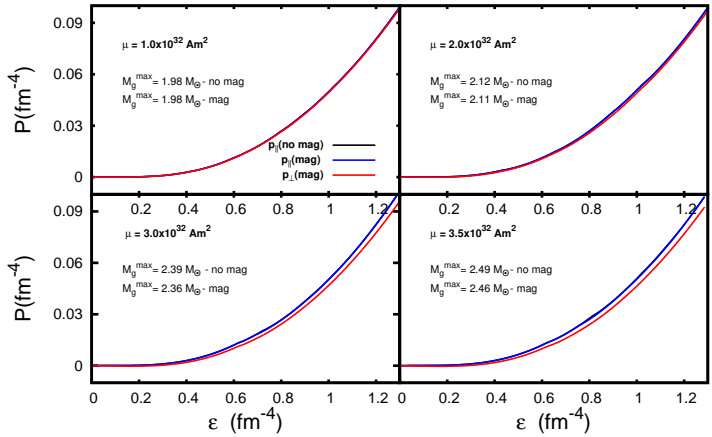}
\vspace*{-.5cm}
\caption{\label{eos_diff_magmom}  Parallel ($\rm{p_{\parallel}}$) and perpendicular ($\rm{p_{\perp}}$) pressure contributions (with respect to B) for different maximum gravitational star masses as shown in Figure~\ref{mr_nocrust_b_magmom}. The perpendicular pressure, $\rm{p_{\perp} = {p_{\parallel}} - mB}$, decreases as the magnetic dipole moment increases.  }
\end{center}
\end {figure}    
    
In Figure~\ref{eos_diff_magmom} we show the EoS$'$s associated with the star presented in Fig.~\ref{mr_nocrust_b_magmom}. They are constructed following the magnetic field profile, which is generated assuming different  magnetic dipole moments. For the pressure in the direction of the magnetic field (parallel) the inclusion or not of the magnetization does not change visibly the magnetic field profile and, therefore,  the pressure in the star (the curves $\rm{p_{\parallel}}$ (no mag) and $\rm{p_{\parallel}}$ (mag)  are overlapped). The parallel pressure does become larger (stiffer EoS) for larger magnetic moments, but such a fact does not reproduce more massive stars.  This stems from the fact that the magnetic field reproduced inside stars in this work is not large enough to turn the EOS much stiffer. In addition, at the center of the star, where the magnetic field is larger,
the baryon number density is reduced (for larger BÕs) and the Haas-van Alphen oscillations make the EOS stiffer and softer at different densities \cite{de1930leiden, Ebert:1999ht, ebert2003quark, ferrari2012chiral}.

 In contrast, the magnetization affects the equation of state contributing to the perpendicular pressure, which is reduced in accordance with the magnetic dipole moment. Due to this, the maximum gravitation masses are also reduced. For example,  the maximum gravitational mass for a dipole magnetic moment of $\mu=3.5\times10^{32}\,\rm{Am^{2}}$ is reduced from 2.49 $\,\rm{M_{\odot}}$ to 2.46 $\,\rm{M_{\odot}}$ when the  magnetization effects are included in the calculation.    
    
\begin{figure}[!t]
\center
\includegraphics[width=0.7\textwidth,angle=-90,scale=0.5]{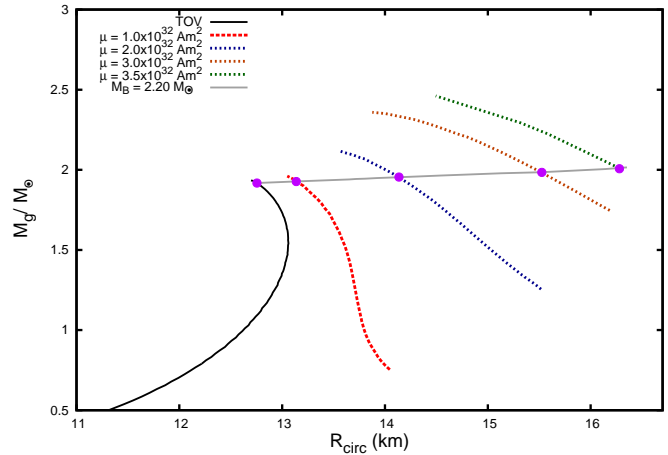}
\caption{\label{mr_nocrust} Mass-radius diagram for non-magnetized and magnetized models. The calculation was done for different fixed magnetic moments $\mu$. The higher the magnetic moment, the higher the magnetic field.  Effects of the magnetic field into the equation of state and the magnetization are also included.  The gray line shows an equilibrium sequence for a fixed baryon mass of $\rm{2.2\,M_{\odot}}$. The full purple circles represent a possible evolution from a highly magnetized neutron star to a non-magnetized and spherical star. }
\end{figure}

From this point on, all the results shown will include, for consistency, the magnetic field effect in the EoS and the magnetization effect. The mass-radius diagram for highly magnetized neutron stars  determined by a constant magnetic dipole moment $\mu$ is presented in Figure \ref{mr_nocrust}. In this figure,  we also show calculations for evolutionary sequences at fixed star baryonic mass of $\rm{M_{B} = 2.2\, M_{\odot}}$.  These lines may represent the transition from a highly magnetized neutron star (a younger star) to a non-magnetized one (an older star).

We have chosen a fixed baryon mass of $\rm{2.2\,M_{\odot}}$ because its evolution line ends almost at the maximum mass for the non-magnetized and spherical configuration. 
Looking at different magnetic dipole moment lines, i.e $\mu = 1.0\times 10^{32}\,\rm{Am^{2}}$, $2.0\times 10^{32}\,\rm{Am^{2}}$, $\mu = 3.0\times 10^{32}\,\rm{Am^{2}}$ and $\mu = 3.5\times 10^{32}\,\rm{Am^{2}}$, one sees that increasing $\mu$ (and therefore the magnetic field) affects the structure of the neutron star in many ways. First, the maximum mass increases, but not so much as in the case of no-fixed baryon number, what had been already raised in Ref.~\cite{Dexheimer:2012mk} using spherical approach. This is an effect of the Lorentz force acting outward and against gravity. For this reason,  the star can support more mass. Second, the circular equatorial radius of the sequence increases and the star becomes much more deformed with respect to the symmetry axis. This deformation is also an effect of the assumption of a poloidal magnetic field, which makes the star more oblate. Calculations including toroidal magnetic field components have shown that magnetized stars become more prolate with respect to the non-magnetized case \cite{Pili:2014npa, Frieben:2012dz, Mastrano:2015rfa}. 

\begin{figure}[!htb]
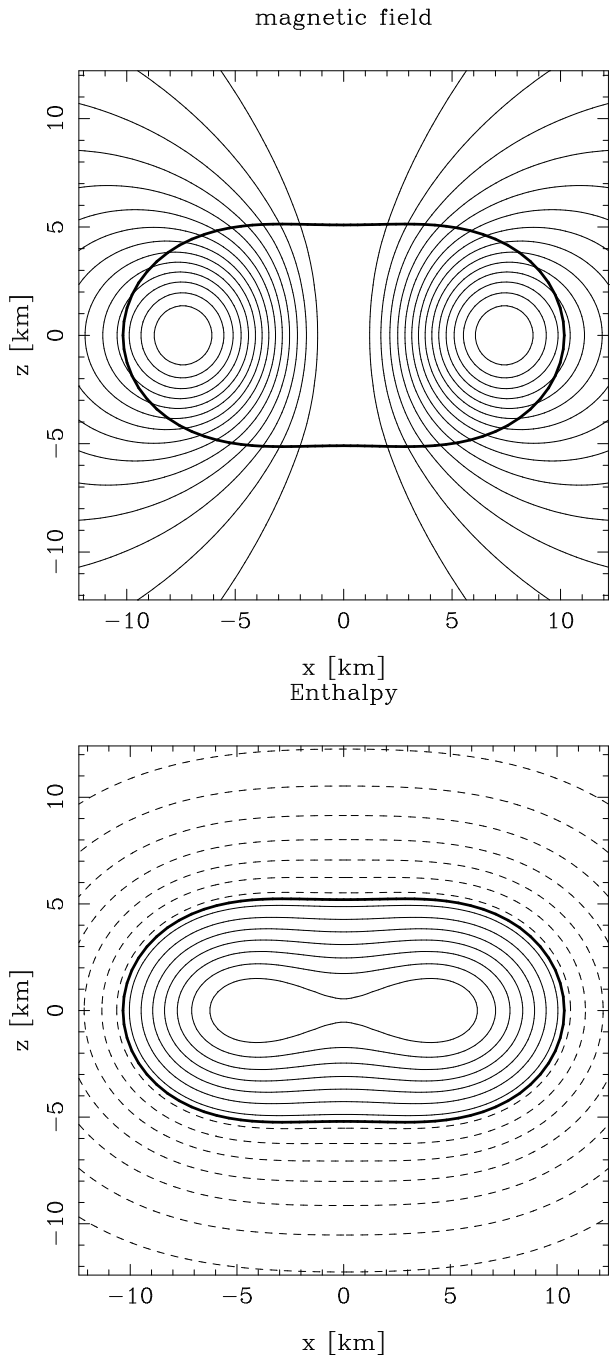

\begin{center}
	\includegraphics[height=8cm, angle=-90]{bfield_static.eps} \quad
	\includegraphics[height=8cm, angle=-90]{enthalpy.eps}
\caption{ Magnetic field surfaces (on the top panel), i.e $A_{\phi}$ iso-contours measured by the Eulerian observer $\mathcal{O}_{0}$ for the chiral EoS. This star is near the maximum equilibrium configuration achieved by the code and the maximum mass for the value $\mu = 3.5\times 10^{32}\rm{\,Am^{2}}$, as shown in Figure \ref{mr_nocrust}.  In the bottom panel, the corresponding enthalpy profile is shown, which corresponds to a central enthalpy of $H_{c}$ = 0.26$\,c^2$ ($n = 0.463\,\,\rm{fm^{-3}})$. The gravitational mass obtained  for the star is $2.46\,\,\rm{M_{\odot}}$ for a central magnetic field of 1.62$\,\times 10^{18}$ G. The ratio between the magnetic pressure and the matter pressure in the center for this star is 0.793.} \label{bfield_static}
\end{center}
\end{figure}

Figure \ref{bfield_static} shows the magnetic field profile and the enthalpy iso-contours for a star with the maximum mass for a magnetization $\mu = 3.5\times10^{32}\,\rm{Am^{2}}$, as can be seen in Figure \ref{mr_nocrust}. This value roughly corresponds to the solution with maximum field configuration achieved with the code.  
\begin{figure}[!t]
\center
\includegraphics[width=0.7\textwidth,angle=-90,scale=0.5]{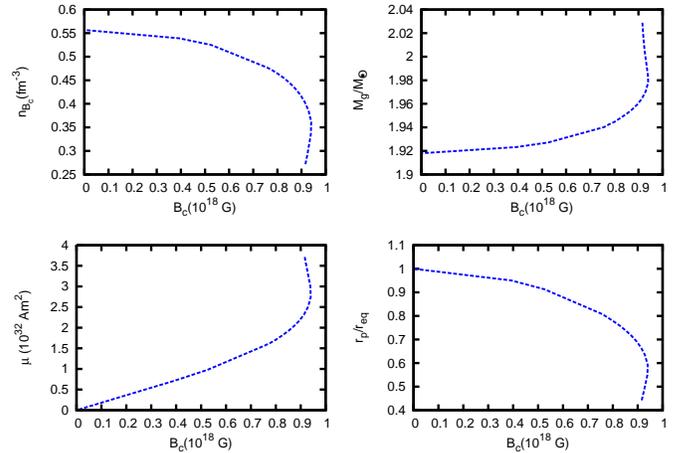}
\caption{\label{globalquantities} Microscopic and macroscopic star quantities, i.e central baryon number, gravitational mass, dipole magnetic moment and ratio between polar and equatorial coordinate radii,  as a function of the central magnetic field.  These curves represent an equilibrium sequence at fixed baryon mass $\rm{ M_{B} = 2.2\, M_{\odot} }$ for different magnetic field intensities. See the line for $\rm{M_{B} = 2.2\, M_{\odot}}$ in Fig. \ref{mr_nocrust}). }
\end{figure}

Now we follow the fixed baryon mass $\rm{M_{B} = 2.2\, M_{\odot}}$ in Figure~\ref{mr_nocrust} and present microscopic and macroscopic properties as a function of the central magnetic field in the stars. As shown in Fig.~\ref{globalquantities}, the central baryon number density decreases with the magnetic field and it has the maximum value at the center only in the static case. 
In the other cases, its maximum is found somewhere inside the star. This is due to the Lorentz force, which is related to the macroscopic currents that create the magnetic field, acting on the matter which has been pushed off-center. This is analogous to the number density reduction in the rotating star case.  
As we will see, this has a huge impact on the particle population of these objects. 
Second, the gravitational mass also increases, as already mentioned in this section. 
Third, as we fixed the baryon mass $M_{B}$, each star in the sequence reproduces different values of the central enthalpy $H_{c}$ and a different magnetic dipole moment $\mu$. 
The latter is free to vary and increases with the magnetic field. Fourth, the ratio between the polar and the equatorial radii increases as the magnetic field increases and, therefore, the star becomes more deformed (oblate).

For all quantities in Figure \ref{globalquantities}, the curves have a qualitatively change in behaviour for a magnetic field strength of $0.9-1.0\times 10^{18}$ G.  At this point, the magnetic force has pushed the matter off-center  and a topological change to a toroidal configuration can take place \cite{cardall2001effects}.  However, our current numerical tools do not enable us to handle toroidal configuration, which gives a limit for the magnetic field strength that we can obtain within this approach. As a consequence, the baryon number density, for example,  will never reach zero at the center of the star. Still, the value of the magnetic field shown in Figure \ref{globalquantities} represents the limit in terms of magnetic field strength for a star at fixed baryon mass of $\rm{M_{B} = 2.2\, M_{\odot}}$. Other configurations, as depicted in Figure \ref{bfield_static}, can reach values higher than $1.0\times10^{18}$ G. 

Note that, in Figure~\ref{globalquantities}, the ratio between the polar and the equatorial radii can reach  50$\%$ for a magnetic field strength of $\sim$ $1\times10^{18}$ G at the center.  Therefore,  one sees that the deviations from spherical symmetry are quite significant and need to be taken into consideration while modelling these highly magnetized objects and a simple TOV solution can not be applied.

\begin{figure}[!t]
\center
\hspace*{-1cm}\includegraphics[width=0.8\textwidth,angle=-90,scale=0.5]{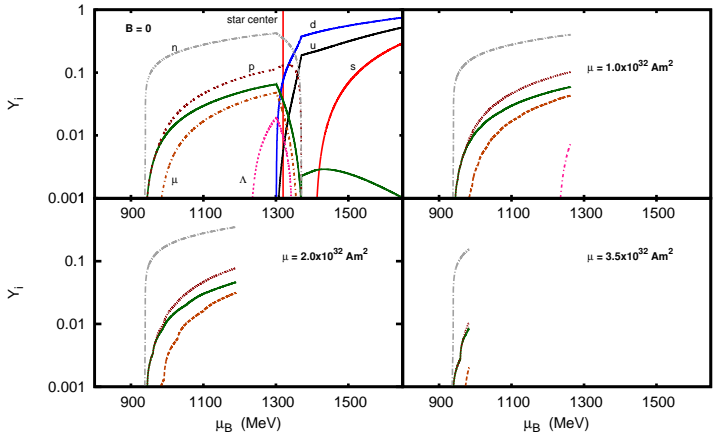}
\caption{\label{population} Stellar particle population as a function of the baryon chemical potential.  All figures represent an equilibrium sequence at fixed baryon mass $\rm{M_{B} = 2.2\, M_{\odot}}$.  As one increases the magnetic field, the particle population changes inside the star.  These stars are represented in Fig. \ref{mr_nocrust} by the full purple circles. For the non-magnetized case (B=0), the vertical red curve represents the chemical potential reached at the center of the star, namely, 1320 MeV.}
\end{figure}

The changes in the global properties of  stars due to the inclusion of the magnetic field into the gravitational equations are remarkable, and in order to study how the microphysics is modified with the magnetic field, we present in Figure \ref{population} the particle population $Y_{i} = \rho_{i}/\rho_{b}$ as a function of the baryon chemical potential $\mu_{B}$ for different values of the magnetic dipole moment $\mu$ for a fixed baryon mass $\rm{M_{B} = 2.2\, M_{\odot}}$ . 
The kinks in the population plot that can be observed  for values equal to or greater than $\mu = 2.0\times 10^{32}\,\rm{Am^{2}}$ are due to the Landau quantization.

For the spherical non-magnetized case, the TOV solution was obtained for a hybrid star  (with mixed phase) composed by the baryon octet, electrons, muons and u, d quarks (no s quark). 
In Figure~\ref{population}, the red vertical line represents the baryon chemical potential reached at the center of the maximum mass star in the non-magnetized case. 
With the inclusion of the magnetic field through the dipole magnetic moment of $\mu = 1.0\times 10^{32}\rm{Am^{2}}$,  the central baryon chemical potential is reduced due to the Lorentz force.  
The new central value for $\mu_{B}$ is below threshold for the creation of quarks, which are, therefore, suppressed.  An even larger effect can be seen in the star for higher values of the magnetic dipole moment $\mu = 2.0\times 10^{32}\,\rm{Am^{2}}$, $\mu = 3.5\times 10^{32}\,\rm{Am^{2}}$, when even the hyperons are suppressed. 
As a result, the properties of these objects such as neutrino emission and consequently the star cooling, are strongly affected by the  magnetic field strength in their interior as already pointed out in Ref.~\cite{Dexheimer:2011pz} for a spherical solution. 

In this way, younger stars that possess strong magnetic fields might go through a phase transition later along their evolution, when their central densities increase enough for the hyperons and quarks to appear. Such phenomena might have observable signatures such as a distinct change in the cooling behaviour as well as  a very different rotational slowing down, reflected in the stellar braking index.

\begin{figure}[!t]
\center
\includegraphics[width=0.7\textwidth,angle=-90,scale=0.5]{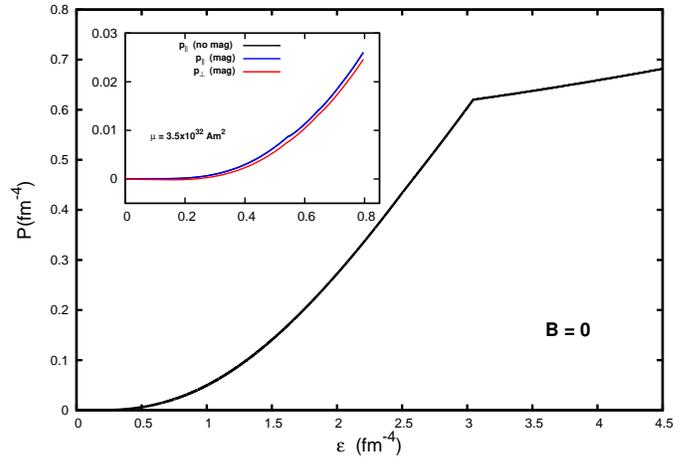}
\caption{\label{eos_mb} Equations of state for a magnetized and non-magnetized star with fixed baryon mass of $\rm{ M_{B} = 2.2\, M_{\odot} }$. In the non-magnetized case,  the equation of state describes a hybrid star with mixed phase (without s quark).  On the other hand, the  magnetized star is described by an equation of state with neutron, protons, electrons and muons.}
\end{figure}

Figure~\ref{eos_mb} depicts the equations of state corresponding to the stars with $\rm{B=0}$ and $\mu = 3.5\times 10^{32}\,\rm{Am^{2}}$ shown in Figure~\ref{population}. For both cases,  the equations of state are shown up to the maximum value of the energy density  reached in center of the star. 
From Figure~\ref{eos_mb}, the suppression of the mixed phase (which in the $\rm{B=0}$ case is in the range $\mathcal{\varepsilon} = [3, 4.5]\, \rm{fm^{-4}}$) is evident. In the case $\mu = 3.5\times 10^{32}\,\rm{Am^{2}}$, as already discussed,
the Lorentz force reduces the energy density in center of the star to a value of $\sim 0.8\, \rm{fm^{-4}}$, i.e. 78$\%$ less that in the non-magnetized case. 
In this same figure, we show also  contributions of the magnetic field and the magnetization to the equation of state. For the values of magnetic fields strength and conditions considered in this work, the former does not  affect the equation of state. 
This can be seen from the components of the parallel pressures, $\rm{p_{\parallel}}\, (no\,\, mag)$ and $\rm{p_{\parallel}}\, (mag)$, which are indistinguishable. 
On the other hand, the magnetization has a small effect on the equation of state, which can be seen from the difference between the parallel and the perpendicular components of the pressures.

 
\section{Conclusion}

We studied highly magnetized neutron stars in a general relativistic framework. 
We solved Einstein-Maxwell equations self-consistently, including a poloidal magnetic field generated by macroscopic currents and taking into account all anisotropies associated with such a field. 
This includes the effects of the magnetic field on the equation of state, including the magnetization. 
In the EoS, we took into consideration the effects of the Landau quantization of the charged particles in the presence of a magnetic field and the anomalous magnetic moment of all hadrons (even the uncharged ones).
 

We found that the leading contribution to the macroscopic properties of stars, like mass and radius, comes from the pure field contribution of the energy-momentum tensor. Still, we found a small contribution when taking into account the magnetic field corrections in the equation of state through the magnetization.  In addition, the breaking of the spherical symmetry and the Lorentz force coming from the macroscopic currents, which generate the magnetic field, change the central density of stars. More precisely, high magnetic fields prevent the appearance of quark and mixed phases in highly magnetized stars.

Assuming that the magnetic field decays over time, stars would not only become less massive and smaller over time, but also go through phase transitions to more exotic phases, thereby potentially significantly modifying their cooling behaviour, which will be studied in detail in future work.

In the future, we are going to include efects due to star
rotation in our calculations, which, for example, can effect the stellar
braking index. Nevertheless, we expect that our findings
are not going to change qualitatively, since rotation is
going to further decrease the central density of the
star



\subsection*{Acknowledgement}
The authors thank to  M. Strickland for fruitful comments and Joachim Frieben and D. Chatterjee for the valuable help with the numerical code. B. Franzon acknowledges support from CNPq/Brazil, DAAD and HGS-HIRe for FAIR. S. Schramm acknowledges support from the HIC for FAIR LOEWE program. The authors wish to acknowledge the "NewCompStar" COST Action MP1304.

 \bibliographystyle{apsrev4-1} 
 \bibliography{bibliografia} 

\begin{thebibliography}{43}%
\makeatletter
\providecommand \@ifxundefined [1]{%
 \@ifx{#1\undefined}
}%
\providecommand \@ifnum [1]{%
 \ifnum #1\expandafter \@firstoftwo
 \else \expandafter \@secondoftwo
 \fi
}%
\providecommand \@ifx [1]{%
 \ifx #1\expandafter \@firstoftwo
 \else \expandafter \@secondoftwo
 \fi
}%
\providecommand \natexlab [1]{#1}%
\providecommand \enquote  [1]{``#1''}%
\providecommand \bibnamefont  [1]{#1}%
\providecommand \bibfnamefont [1]{#1}%
\providecommand \citenamefont [1]{#1}%
\providecommand \href@noop [0]{\@secondoftwo}%
\providecommand \href [0]{\begingroup \@sanitize@url \@href}%
\providecommand \@href[1]{\@@startlink{#1}\@@href}%
\providecommand \@@href[1]{\endgroup#1\@@endlink}%
\providecommand \@sanitize@url [0]{\catcode `\\12\catcode `\$12\catcode
  `\&12\catcode `\#12\catcode `\^12\catcode `\_12\catcode `\%12\relax}%
\providecommand \@@startlink[1]{}%
\providecommand \@@endlink[0]{}%
\providecommand \url  [0]{\begingroup\@sanitize@url \@url }%
\providecommand \@url [1]{\endgroup\@href {#1}{\urlprefix }}%
\providecommand \urlprefix  [0]{URL }%
\providecommand \Eprint [0]{\href }%
\providecommand \doibase [0]{http://dx.doi.org/}%
\providecommand \selectlanguage [0]{\@gobble}%
\providecommand \bibinfo  [0]{\@secondoftwo}%
\providecommand \bibfield  [0]{\@secondoftwo}%
\providecommand \translation [1]{[#1]}%
\providecommand \BibitemOpen [0]{}%
\providecommand \bibitemStop [0]{}%
\providecommand \bibitemNoStop [0]{.\EOS\space}%
\providecommand \EOS [0]{\spacefactor3000\relax}%
\providecommand \BibitemShut  [1]{\csname bibitem#1\endcsname}%
\let\auto@bib@innerbib\@empty
\bibitem [{\citenamefont {Duncan}\ and\ \citenamefont
  {Thompson}(1992)}]{Duncan:1992hi}%
  \BibitemOpen
  \bibfield  {author} {\bibinfo {author} {\bibfnamefont {R.~C.}\ \bibnamefont
  {Duncan}}\ and\ \bibinfo {author} {\bibfnamefont {C.}~\bibnamefont
  {Thompson}},\ }\href {\doibase 10.1086/186413} {\bibfield  {journal}
  {\bibinfo  {journal} {Astrophys. J.}\ }\textbf {\bibinfo {volume} {392}},\
  \bibinfo {pages} {L9} (\bibinfo {year} {1992})}\BibitemShut {NoStop}%
\bibitem [{\citenamefont {Thompson}\ and\ \citenamefont
  {Duncan}(1993)}]{Thompson:1993hn}%
  \BibitemOpen
  \bibfield  {author} {\bibinfo {author} {\bibfnamefont {C.}~\bibnamefont
  {Thompson}}\ and\ \bibinfo {author} {\bibfnamefont {R.~C.}\ \bibnamefont
  {Duncan}},\ }\href {\doibase 10.1086/172580} {\bibfield  {journal} {\bibinfo
  {journal} {Astrophys. J.}\ }\textbf {\bibinfo {volume} {408}},\ \bibinfo
  {pages} {194} (\bibinfo {year} {1993})}\BibitemShut {NoStop}%
\bibitem [{\citenamefont {Thompson}\ and\ \citenamefont
  {Duncan}(1996)}]{Thompson:1996pe}%
  \BibitemOpen
  \bibfield  {author} {\bibinfo {author} {\bibfnamefont {C.}~\bibnamefont
  {Thompson}}\ and\ \bibinfo {author} {\bibfnamefont {R.~C.}\ \bibnamefont
  {Duncan}},\ }\href {\doibase 10.1086/178147} {\bibfield  {journal} {\bibinfo
  {journal} {Astrophys. J.}\ }\textbf {\bibinfo {volume} {473}},\ \bibinfo
  {pages} {322} (\bibinfo {year} {1996})}\BibitemShut {NoStop}%
\bibitem [{\citenamefont {Paczynski}(1992)}]{paczynski1992gb}%
  \BibitemOpen
  \bibfield  {author} {\bibinfo {author} {\bibfnamefont {B.}~\bibnamefont
  {Paczynski}},\ }\href@noop {} {\bibfield  {journal} {\bibinfo  {journal}
  {Acta Astronomica}\ }\textbf {\bibinfo {volume} {42}},\ \bibinfo {pages}
  {145} (\bibinfo {year} {1992})}\BibitemShut {NoStop}%
\bibitem [{\citenamefont {Melatos}(1999)}]{melatos1999bumpy}%
  \BibitemOpen
  \bibfield  {author} {\bibinfo {author} {\bibfnamefont {A.}~\bibnamefont
  {Melatos}},\ }\href@noop {} {\bibfield  {journal} {\bibinfo  {journal} {The
  Astrophysical Journal Letters}\ }\textbf {\bibinfo {volume} {519}},\ \bibinfo
  {pages} {L77} (\bibinfo {year} {1999})}\BibitemShut {NoStop}%
\bibitem [{\citenamefont {Makishima}\ \emph {et~al.}(2014)\citenamefont
  {Makishima}, \citenamefont {Enoto}, \citenamefont {Hiraga}, \citenamefont
  {Nakano}, \citenamefont {Nakazawa}, \citenamefont {Sakurai}, \citenamefont
  {Sasano},\ and\ \citenamefont {Murakami}}]{Makishima:2014dua}%
  \BibitemOpen
  \bibfield  {author} {\bibinfo {author} {\bibfnamefont {K.}~\bibnamefont
  {Makishima}}, \bibinfo {author} {\bibfnamefont {T.}~\bibnamefont {Enoto}},
  \bibinfo {author} {\bibfnamefont {J.~S.}\ \bibnamefont {Hiraga}}, \bibinfo
  {author} {\bibfnamefont {T.}~\bibnamefont {Nakano}}, \bibinfo {author}
  {\bibfnamefont {K.}~\bibnamefont {Nakazawa}}, \bibinfo {author}
  {\bibfnamefont {S.}~\bibnamefont {Sakurai}}, \bibinfo {author} {\bibfnamefont
  {M.}~\bibnamefont {Sasano}}, \ and\ \bibinfo {author} {\bibfnamefont
  {H.}~\bibnamefont {Murakami}},\ }\href {\doibase
  10.1103/PhysRevLett.112.171102} {\bibfield  {journal} {\bibinfo  {journal}
  {Phys. Rev. Lett.}\ }\textbf {\bibinfo {volume} {112}},\ \bibinfo {pages}
  {171102} (\bibinfo {year} {2014})},\ \Eprint {http://arxiv.org/abs/1404.3705}
  {arXiv:1404.3705 [astro-ph.HE]} \BibitemShut {NoStop}%
\bibitem [{\citenamefont {Ferrer}\ \emph {et~al.}(2010)\citenamefont {Ferrer},
  \citenamefont {de~La~Incera}, \citenamefont {Keith}, \citenamefont
  {Portillo},\ and\ \citenamefont {Springsteen}}]{ferrer2010equation}%
  \BibitemOpen
  \bibfield  {author} {\bibinfo {author} {\bibfnamefont {E.~J.}\ \bibnamefont
  {Ferrer}}, \bibinfo {author} {\bibfnamefont {V.}~\bibnamefont
  {de~La~Incera}}, \bibinfo {author} {\bibfnamefont {J.~P.}\ \bibnamefont
  {Keith}}, \bibinfo {author} {\bibfnamefont {I.}~\bibnamefont {Portillo}}, \
  and\ \bibinfo {author} {\bibfnamefont {P.~L.}\ \bibnamefont {Springsteen}},\
  }\href@noop {} {\bibfield  {journal} {\bibinfo  {journal} {Physical Review
  C}\ }\textbf {\bibinfo {volume} {82}},\ \bibinfo {pages} {065802} (\bibinfo
  {year} {2010})}\BibitemShut {NoStop}%
\bibitem [{\citenamefont {Lai}\ and\ \citenamefont
  {Shapiro}(1991)}]{lai1991cold}%
  \BibitemOpen
  \bibfield  {author} {\bibinfo {author} {\bibfnamefont {D.}~\bibnamefont
  {Lai}}\ and\ \bibinfo {author} {\bibfnamefont {S.~L.}\ \bibnamefont
  {Shapiro}},\ }\href@noop {} {\bibfield  {journal} {\bibinfo  {journal} {The
  Astrophysical Journal}\ }\textbf {\bibinfo {volume} {383}},\ \bibinfo {pages}
  {745} (\bibinfo {year} {1991})}\BibitemShut {NoStop}%
\bibitem [{\citenamefont {Fushiki}\ \emph {et~al.}(1989)\citenamefont
  {Fushiki}, \citenamefont {Gudmundsson},\ and\ \citenamefont
  {Pethick}}]{fushiki1989surface}%
  \BibitemOpen
  \bibfield  {author} {\bibinfo {author} {\bibfnamefont {I.}~\bibnamefont
  {Fushiki}}, \bibinfo {author} {\bibfnamefont {E.}~\bibnamefont
  {Gudmundsson}}, \ and\ \bibinfo {author} {\bibfnamefont {C.}~\bibnamefont
  {Pethick}},\ }\href@noop {} {\bibfield  {journal} {\bibinfo  {journal} {The
  Astrophysical Journal}\ }\textbf {\bibinfo {volume} {342}},\ \bibinfo {pages}
  {958} (\bibinfo {year} {1989})}\BibitemShut {NoStop}%
\bibitem [{\citenamefont {Cardall}\ \emph {et~al.}(2001)\citenamefont
  {Cardall}, \citenamefont {Prakash},\ and\ \citenamefont
  {Lattimer}}]{cardall2001effects}%
  \BibitemOpen
  \bibfield  {author} {\bibinfo {author} {\bibfnamefont {C.~Y.}\ \bibnamefont
  {Cardall}}, \bibinfo {author} {\bibfnamefont {M.}~\bibnamefont {Prakash}}, \
  and\ \bibinfo {author} {\bibfnamefont {J.~M.}\ \bibnamefont {Lattimer}},\
  }\href@noop {} {\bibfield  {journal} {\bibinfo  {journal} {The Astrophysical
  Journal}\ }\textbf {\bibinfo {volume} {554}},\ \bibinfo {pages} {322}
  (\bibinfo {year} {2001})}\BibitemShut {NoStop}%
\bibitem [{\citenamefont {L.}(1964)}]{woltjer}%
  \BibitemOpen
  \bibfield  {author} {\bibinfo {author} {\bibfnamefont {W.}~\bibnamefont
  {L.}},\ }\href@noop {} {\bibfield  {journal} {\bibinfo  {journal} {The
  Astrophysical Journal}\ }\textbf {\bibinfo {volume} {140}},\ \bibinfo {pages}
  {1309} (\bibinfo {year} {1964})}\BibitemShut {NoStop}%
\bibitem [{\citenamefont {Tatsumi}(2000)}]{tatsumi2000ferromagnetism}%
  \BibitemOpen
  \bibfield  {author} {\bibinfo {author} {\bibfnamefont {T.}~\bibnamefont
  {Tatsumi}},\ }\href@noop {} {\bibfield  {journal} {\bibinfo  {journal}
  {Physics Letters B}\ }\textbf {\bibinfo {volume} {489}},\ \bibinfo {pages}
  {280} (\bibinfo {year} {2000})}\BibitemShut {NoStop}%
\bibitem [{\citenamefont {Vink}\ and\ \citenamefont
  {Kuiper}(2006)}]{vink2006supernova}%
  \BibitemOpen
  \bibfield  {author} {\bibinfo {author} {\bibfnamefont {J.}~\bibnamefont
  {Vink}}\ and\ \bibinfo {author} {\bibfnamefont {L.}~\bibnamefont {Kuiper}},\
  }\href@noop {} {\bibfield  {journal} {\bibinfo  {journal} {Monthly Notices of
  the Royal Astronomical Society: Letters}\ }\textbf {\bibinfo {volume}
  {370}},\ \bibinfo {pages} {L14} (\bibinfo {year} {2006})}\BibitemShut
  {NoStop}%
\bibitem [{\citenamefont {Tolman}(1939)}]{tolman1939static}%
  \BibitemOpen
  \bibfield  {author} {\bibinfo {author} {\bibfnamefont {R.~C.}\ \bibnamefont
  {Tolman}},\ }\href@noop {} {\bibfield  {journal} {\bibinfo  {journal}
  {Physical Review}\ }\textbf {\bibinfo {volume} {55}},\ \bibinfo {pages} {364}
  (\bibinfo {year} {1939})}\BibitemShut {NoStop}%
\bibitem [{\citenamefont {Oppenheimer}\ and\ \citenamefont
  {Volkoff}(1939)}]{oppenheimer1939massive}%
  \BibitemOpen
  \bibfield  {author} {\bibinfo {author} {\bibfnamefont {J.~R.}\ \bibnamefont
  {Oppenheimer}}\ and\ \bibinfo {author} {\bibfnamefont {G.~M.}\ \bibnamefont
  {Volkoff}},\ }\href@noop {} {\bibfield  {journal} {\bibinfo  {journal}
  {Physical Review}\ }\textbf {\bibinfo {volume} {55}},\ \bibinfo {pages} {374}
  (\bibinfo {year} {1939})}\BibitemShut {NoStop}%
\bibitem [{\citenamefont {Bonazzola}\ \emph {et~al.}(1993)\citenamefont
  {Bonazzola}, \citenamefont {Gourgoulhon}, \citenamefont {Salgado},\ and\
  \citenamefont {Marck}}]{Bonazzola:1993zz}%
  \BibitemOpen
  \bibfield  {author} {\bibinfo {author} {\bibfnamefont {S.}~\bibnamefont
  {Bonazzola}}, \bibinfo {author} {\bibfnamefont {E.}~\bibnamefont
  {Gourgoulhon}}, \bibinfo {author} {\bibfnamefont {M.}~\bibnamefont
  {Salgado}}, \ and\ \bibinfo {author} {\bibfnamefont {J.~A. t. A. r. r. b. A.
  n. n. a. f. e.~s.}\ \bibnamefont {Marck}},\ }\href@noop {} {\bibfield
  {journal} {\bibinfo  {journal} {Astron. Astrophys.}\ }\textbf {\bibinfo
  {volume} {278}},\ \bibinfo {pages} {421} (\bibinfo {year}
  {1993})}\BibitemShut {NoStop}%
\bibitem [{\citenamefont {Chatterjee}\ \emph {et~al.}(2015)\citenamefont
  {Chatterjee}, \citenamefont {Elghozi}, \citenamefont {Novak},\ and\
  \citenamefont {Oertel}}]{Chatterjee:2014qsa}%
  \BibitemOpen
  \bibfield  {author} {\bibinfo {author} {\bibfnamefont {D.}~\bibnamefont
  {Chatterjee}}, \bibinfo {author} {\bibfnamefont {T.}~\bibnamefont {Elghozi}},
  \bibinfo {author} {\bibfnamefont {J.}~\bibnamefont {Novak}}, \ and\ \bibinfo
  {author} {\bibfnamefont {M.}~\bibnamefont {Oertel}},\ }\href {\doibase
  10.1093/mnras/stu2706} {\bibfield  {journal} {\bibinfo  {journal} {Mon. Not.
  Roy. Astron. Soc.}\ }\textbf {\bibinfo {volume} {447}},\ \bibinfo {pages}
  {3785} (\bibinfo {year} {2015})},\ \Eprint {http://arxiv.org/abs/1410.6332}
  {arXiv:1410.6332 [astro-ph.HE]} \BibitemShut {NoStop}%
\bibitem [{\citenamefont {Noronha}\ and\ \citenamefont
  {Shovkovy}(2007)}]{noronha2007color}%
  \BibitemOpen
  \bibfield  {author} {\bibinfo {author} {\bibfnamefont {J.~L.}\ \bibnamefont
  {Noronha}}\ and\ \bibinfo {author} {\bibfnamefont {I.~A.}\ \bibnamefont
  {Shovkovy}},\ }\href@noop {} {\bibfield  {journal} {\bibinfo  {journal}
  {Physical Review D}\ }\textbf {\bibinfo {volume} {76}},\ \bibinfo {pages}
  {105030} (\bibinfo {year} {2007})}\BibitemShut {NoStop}%
\bibitem [{\citenamefont {Bocquet}\ \emph {et~al.}(1995)\citenamefont
  {Bocquet}, \citenamefont {Bonazzola}, \citenamefont {Gourgoulhon},\ and\
  \citenamefont {Novak}}]{Bocquet:1995je}%
  \BibitemOpen
  \bibfield  {author} {\bibinfo {author} {\bibfnamefont {M.}~\bibnamefont
  {Bocquet}}, \bibinfo {author} {\bibfnamefont {S.}~\bibnamefont {Bonazzola}},
  \bibinfo {author} {\bibfnamefont {E.}~\bibnamefont {Gourgoulhon}}, \ and\
  \bibinfo {author} {\bibfnamefont {J.}~\bibnamefont {Novak}},\ }\href@noop {}
  {\bibfield  {journal} {\bibinfo  {journal} {Astron. Astrophys.}\ }\textbf
  {\bibinfo {volume} {301}},\ \bibinfo {pages} {757} (\bibinfo {year}
  {1995})},\ \Eprint {http://arxiv.org/abs/gr-qc/9503044} {arXiv:gr-qc/9503044
  [gr-qc]} \BibitemShut {NoStop}%
\bibitem [{\citenamefont {Gourgoulhon}(2012)}]{gourgoulhon20123+}%
  \BibitemOpen
  \bibfield  {author} {\bibinfo {author} {\bibfnamefont {E.}~\bibnamefont
  {Gourgoulhon}},\ }\href@noop {} {\emph {\bibinfo {title} {3+ 1 formalism in
  general relativity: bases of numerical relativity}}},\ Vol.\ \bibinfo
  {volume} {846}\ (\bibinfo  {publisher} {Springer Science \& Business Media},\
  \bibinfo {year} {2012})\BibitemShut {NoStop}%
\bibitem [{\citenamefont {Lichnerowicz}\ \emph {et~al.}(1967)\citenamefont
  {Lichnerowicz}, \citenamefont {for Advanced~Studies},\ and\ \citenamefont
  {Series}}]{lichnerowicz1967relativistic}%
  \BibitemOpen
  \bibfield  {author} {\bibinfo {author} {\bibfnamefont {A.}~\bibnamefont
  {Lichnerowicz}}, \bibinfo {author} {\bibfnamefont {S.~C.}\ \bibnamefont {for
  Advanced~Studies}}, \ and\ \bibinfo {author} {\bibfnamefont {T.~M. P.~M.}\
  \bibnamefont {Series}},\ }\href@noop {} {\emph {\bibinfo {title}
  {Relativistic hydrodynamics and magnetohydrodynamics}}},\ Vol.~\bibinfo
  {volume} {35}\ (\bibinfo  {publisher} {WA Benjamin New York},\ \bibinfo
  {year} {1967})\BibitemShut {NoStop}%
\bibitem [{\citenamefont {Hempel}\ \emph {et~al.}(2013)\citenamefont {Hempel},
  \citenamefont {Dexheimer}, \citenamefont {Schramm},\ and\ \citenamefont
  {Iosilevskiy}}]{PhysRevC.88.014906}%
  \BibitemOpen
  \bibfield  {author} {\bibinfo {author} {\bibfnamefont {M.}~\bibnamefont
  {Hempel}}, \bibinfo {author} {\bibfnamefont {V.}~\bibnamefont {Dexheimer}},
  \bibinfo {author} {\bibfnamefont {S.}~\bibnamefont {Schramm}}, \ and\
  \bibinfo {author} {\bibfnamefont {I.}~\bibnamefont {Iosilevskiy}},\ }\href
  {\doibase 10.1103/PhysRevC.88.014906} {\bibfield  {journal} {\bibinfo
  {journal} {Phys. Rev. C}\ }\textbf {\bibinfo {volume} {88}},\ \bibinfo
  {pages} {014906} (\bibinfo {year} {2013})}\BibitemShut {NoStop}%
\bibitem [{\citenamefont {Papazoglou}\ \emph {et~al.}(1999)\citenamefont
  {Papazoglou}, \citenamefont {Zschiesche}, \citenamefont {Schramm},
  \citenamefont {Schaffner-Bielich}, \citenamefont {Stoecker} \emph
  {et~al.}}]{Papazoglou:1998vr}%
  \BibitemOpen
  \bibfield  {author} {\bibinfo {author} {\bibfnamefont {P.}~\bibnamefont
  {Papazoglou}}, \bibinfo {author} {\bibfnamefont {D.}~\bibnamefont
  {Zschiesche}}, \bibinfo {author} {\bibfnamefont {S.}~\bibnamefont {Schramm}},
  \bibinfo {author} {\bibfnamefont {J.}~\bibnamefont {Schaffner-Bielich}},
  \bibinfo {author} {\bibfnamefont {H.}~\bibnamefont {Stoecker}},  \emph
  {et~al.},\ }\href {\doibase 10.1103/PhysRevC.59.411} {\bibfield  {journal}
  {\bibinfo  {journal} {Phys.Rev.}\ }\textbf {\bibinfo {volume} {C59}},\
  \bibinfo {pages} {411} (\bibinfo {year} {1999})},\ \Eprint
  {http://arxiv.org/abs/nucl-th/9806087} {arXiv:nucl-th/9806087 [nucl-th]}
  \BibitemShut {NoStop}%
\bibitem [{\citenamefont {Dexheimer}\ and\ \citenamefont
  {Schramm}(2008)}]{Dexheimer:2008ax}%
  \BibitemOpen
  \bibfield  {author} {\bibinfo {author} {\bibfnamefont {V.}~\bibnamefont
  {Dexheimer}}\ and\ \bibinfo {author} {\bibfnamefont {S.}~\bibnamefont
  {Schramm}},\ }\href {\doibase 10.1086/589735} {\bibfield  {journal} {\bibinfo
   {journal} {Astrophys.J.}\ }\textbf {\bibinfo {volume} {683}},\ \bibinfo
  {pages} {943} (\bibinfo {year} {2008})},\ \Eprint
  {http://arxiv.org/abs/0802.1999} {arXiv:0802.1999 [astro-ph]} \BibitemShut
  {NoStop}%
\bibitem [{\citenamefont {Dexheimer}\ and\ \citenamefont
  {Schramm}(2010)}]{Dexheimer:2009hi}%
  \BibitemOpen
  \bibfield  {author} {\bibinfo {author} {\bibfnamefont {V.}~\bibnamefont
  {Dexheimer}}\ and\ \bibinfo {author} {\bibfnamefont {S.}~\bibnamefont
  {Schramm}},\ }\href {\doibase 10.1103/PhysRevC.81.045201} {\bibfield
  {journal} {\bibinfo  {journal} {Phys.Rev.}\ }\textbf {\bibinfo {volume}
  {C81}},\ \bibinfo {pages} {045201} (\bibinfo {year} {2010})},\ \Eprint
  {http://arxiv.org/abs/0901.1748} {arXiv:0901.1748 [astro-ph.SR]} \BibitemShut
  {NoStop}%
\bibitem [{\citenamefont {Fukushima}(2004)}]{Fukushima:2003fw}%
  \BibitemOpen
  \bibfield  {author} {\bibinfo {author} {\bibfnamefont {K.}~\bibnamefont
  {Fukushima}},\ }\href {\doibase 10.1016/j.physletb.2004.04.027} {\bibfield
  {journal} {\bibinfo  {journal} {Phys.Lett.}\ }\textbf {\bibinfo {volume}
  {B591}},\ \bibinfo {pages} {277} (\bibinfo {year} {2004})},\ \Eprint
  {http://arxiv.org/abs/hep-ph/0310121} {arXiv:hep-ph/0310121 [hep-ph]}
  \BibitemShut {NoStop}%
\bibitem [{\citenamefont {Weinberg}(1990)}]{Weinberg:1990xm}%
  \BibitemOpen
  \bibfield  {author} {\bibinfo {author} {\bibfnamefont {S.}~\bibnamefont
  {Weinberg}},\ }\href {\doibase 10.1103/PhysRevLett.65.1181} {\bibfield
  {journal} {\bibinfo  {journal} {Phys.Rev.Lett.}\ }\textbf {\bibinfo {volume}
  {65}},\ \bibinfo {pages} {1181} (\bibinfo {year} {1990})}\BibitemShut
  {NoStop}%
\bibitem [{\citenamefont {Ferrer}\ \emph {et~al.}(2015)\citenamefont {Ferrer},
  \citenamefont {de~la Incera}, \citenamefont {Paret}, \citenamefont
  {Mart\'{i}nez},\ and\ \citenamefont {Sanchez}}]{PhysRevD.91.085041}%
  \BibitemOpen
  \bibfield  {author} {\bibinfo {author} {\bibfnamefont {E.~J.}\ \bibnamefont
  {Ferrer}}, \bibinfo {author} {\bibfnamefont {V.}~\bibnamefont {de~la
  Incera}}, \bibinfo {author} {\bibfnamefont {D.~M.}\ \bibnamefont {Paret}},
  \bibinfo {author} {\bibfnamefont {A.~P.}\ \bibnamefont {Mart\'{i}nez}}, \
  and\ \bibinfo {author} {\bibfnamefont {A.}~\bibnamefont {Sanchez}},\ }\href
  {\doibase 10.1103/PhysRevD.91.085041} {\bibfield  {journal} {\bibinfo
  {journal} {Phys. Rev. D}\ }\textbf {\bibinfo {volume} {91}},\ \bibinfo
  {pages} {085041} (\bibinfo {year} {2015})}\BibitemShut {NoStop}%
\bibitem [{\citenamefont {Chakrabarty}(1996)}]{Chakrabarty:1996te}%
  \BibitemOpen
  \bibfield  {author} {\bibinfo {author} {\bibfnamefont {S.}~\bibnamefont
  {Chakrabarty}},\ }\href@noop {} {\bibfield  {journal} {\bibinfo  {journal}
  {Phys. Rev.}\ }\textbf {\bibinfo {volume} {D54}},\ \bibinfo {pages} {1306}
  (\bibinfo {year} {1996})}\BibitemShut {NoStop}%
\bibitem [{\citenamefont {Suh}\ \emph {et~al.}(2001)\citenamefont {Suh},
  \citenamefont {Mathews},\ and\ \citenamefont {Weber}}]{Suh:2001tr}%
  \BibitemOpen
  \bibfield  {author} {\bibinfo {author} {\bibfnamefont {I.-S.}\ \bibnamefont
  {Suh}}, \bibinfo {author} {\bibfnamefont {G.}~\bibnamefont {Mathews}}, \ and\
  \bibinfo {author} {\bibfnamefont {F.}~\bibnamefont {Weber}},\ }\href@noop {}
  {\bibfield  {journal} {\bibinfo  {journal} {Phys.Rev.D}\ } (\bibinfo {year}
  {2001})}\BibitemShut {NoStop}%
\bibitem [{\citenamefont {Perez~Martinez}\ \emph {et~al.}(2005)\citenamefont
  {Perez~Martinez}, \citenamefont {Perez~Rojas}, \citenamefont
  {Mosquera~Cuesta}, \citenamefont {Boligan},\ and\ \citenamefont
  {Orsaria}}]{PerezMartinez:2005av}%
  \BibitemOpen
  \bibfield  {author} {\bibinfo {author} {\bibfnamefont {A.}~\bibnamefont
  {Perez~Martinez}}, \bibinfo {author} {\bibfnamefont {H.}~\bibnamefont
  {Perez~Rojas}}, \bibinfo {author} {\bibfnamefont {H.~J.}\ \bibnamefont
  {Mosquera~Cuesta}}, \bibinfo {author} {\bibfnamefont {M.}~\bibnamefont
  {Boligan}}, \ and\ \bibinfo {author} {\bibfnamefont {M.~G.}\ \bibnamefont
  {Orsaria}},\ }\href@noop {} {\bibfield  {journal} {\bibinfo  {journal} {Int.
  J. Mod. Phys.}\ }\textbf {\bibinfo {volume} {D14}},\ \bibinfo {pages} {1959}
  (\bibinfo {year} {2005})}\BibitemShut {NoStop}%
\bibitem [{\citenamefont {Felipe}\ \emph {et~al.}(2008)\citenamefont {Felipe},
  \citenamefont {Martinez}, \citenamefont {Rojas},\ and\ \citenamefont
  {Orsaria}}]{Felipe:2007vb}%
  \BibitemOpen
  \bibfield  {author} {\bibinfo {author} {\bibfnamefont {R.~G.}\ \bibnamefont
  {Felipe}}, \bibinfo {author} {\bibfnamefont {A.~P.}\ \bibnamefont
  {Martinez}}, \bibinfo {author} {\bibfnamefont {H.~P.}\ \bibnamefont {Rojas}},
  \ and\ \bibinfo {author} {\bibfnamefont {M.}~\bibnamefont {Orsaria}},\
  }\href@noop {} {\bibfield  {journal} {\bibinfo  {journal} {Phys. Rev.}\
  }\textbf {\bibinfo {volume} {C77}},\ \bibinfo {pages} {015807} (\bibinfo
  {year} {2008})}\BibitemShut {NoStop}%
\bibitem [{\citenamefont {Dexheimer}\ \emph {et~al.}(2012)\citenamefont
  {Dexheimer}, \citenamefont {Negreiros},\ and\ \citenamefont
  {Schramm}}]{Dexheimer:2011pz}%
  \BibitemOpen
  \bibfield  {author} {\bibinfo {author} {\bibfnamefont {V.}~\bibnamefont
  {Dexheimer}}, \bibinfo {author} {\bibfnamefont {R.}~\bibnamefont
  {Negreiros}}, \ and\ \bibinfo {author} {\bibfnamefont {S.}~\bibnamefont
  {Schramm}},\ }\href {\doibase 10.1140/epja/i2012-12189-y} {\bibfield
  {journal} {\bibinfo  {journal} {Eur.Phys.J.}\ }\textbf {\bibinfo {volume}
  {A48}},\ \bibinfo {pages} {189} (\bibinfo {year} {2012})},\ \Eprint
  {http://arxiv.org/abs/1108.4479} {arXiv:1108.4479 [astro-ph.HE]} \BibitemShut
  {NoStop}%
\bibitem [{\citenamefont {Dexheimer}\ \emph {et~al.}(2013)\citenamefont
  {Dexheimer}, \citenamefont {Negreiros}, \citenamefont {Schramm},\ and\
  \citenamefont {Hempel}}]{Dexheimer:2012qk}%
  \BibitemOpen
  \bibfield  {author} {\bibinfo {author} {\bibfnamefont {V.}~\bibnamefont
  {Dexheimer}}, \bibinfo {author} {\bibfnamefont {R.}~\bibnamefont
  {Negreiros}}, \bibinfo {author} {\bibfnamefont {S.}~\bibnamefont {Schramm}},
  \ and\ \bibinfo {author} {\bibfnamefont {M.}~\bibnamefont {Hempel}},\ }\href
  {\doibase 10.1063/1.4795968} {\bibfield  {journal} {\bibinfo  {journal} {AIP
  Conf.Proc.}\ }\textbf {\bibinfo {volume} {1520}},\ \bibinfo {pages} {264}
  (\bibinfo {year} {2013})},\ \Eprint {http://arxiv.org/abs/1208.1320}
  {arXiv:1208.1320 [astro-ph.HE]} \BibitemShut {NoStop}%
\bibitem [{\citenamefont {Strickland}\ \emph {et~al.}(2012)\citenamefont
  {Strickland}, \citenamefont {Dexheimer},\ and\ \citenamefont
  {Menezes}}]{Strickland:2012vu}%
  \BibitemOpen
  \bibfield  {author} {\bibinfo {author} {\bibfnamefont {M.}~\bibnamefont
  {Strickland}}, \bibinfo {author} {\bibfnamefont {V.}~\bibnamefont
  {Dexheimer}}, \ and\ \bibinfo {author} {\bibfnamefont {D.~P.}\ \bibnamefont
  {Menezes}},\ }\href {\doibase 10.1103/PhysRevD.86.125032} {\bibfield
  {journal} {\bibinfo  {journal} {Phys. Rev.}\ }\textbf {\bibinfo {volume}
  {D86}},\ \bibinfo {pages} {125032} (\bibinfo {year} {2012})},\ \Eprint
  {http://arxiv.org/abs/1209.3276} {arXiv:1209.3276 [nucl-th]} \BibitemShut
  {NoStop}%
\bibitem [{\citenamefont {de~Haas}\ and\ \citenamefont {van
  Alphen}(1930)}]{de1930leiden}%
  \BibitemOpen
  \bibfield  {author} {\bibinfo {author} {\bibfnamefont {W.}~\bibnamefont
  {de~Haas}}\ and\ \bibinfo {author} {\bibfnamefont {P.}~\bibnamefont {van
  Alphen}},\ }in\ \href@noop {} {\emph {\bibinfo {booktitle} {Proc. R. Acad.
  Sci. Amsterdam}}},\ Vol.~\bibinfo {volume} {33}\ (\bibinfo {year} {1930})\
  p.\ \bibinfo {pages} {1106}\BibitemShut {NoStop}%
\bibitem [{\citenamefont {Ebert}\ \emph {et~al.}(2000)\citenamefont {Ebert},
  \citenamefont {Klimenko}, \citenamefont {Vdovichenko},\ and\ \citenamefont
  {Vshivtsev}}]{Ebert:1999ht}%
  \BibitemOpen
  \bibfield  {author} {\bibinfo {author} {\bibfnamefont {D.}~\bibnamefont
  {Ebert}}, \bibinfo {author} {\bibfnamefont {K.~G.}\ \bibnamefont {Klimenko}},
  \bibinfo {author} {\bibfnamefont {M.~A.}\ \bibnamefont {Vdovichenko}}, \ and\
  \bibinfo {author} {\bibfnamefont {A.~S.}\ \bibnamefont {Vshivtsev}},\ }\href
  {\doibase 10.1103/PhysRevD.61.025005} {\bibfield  {journal} {\bibinfo
  {journal} {Phys. Rev.}\ }\textbf {\bibinfo {volume} {D61}},\ \bibinfo {pages}
  {025005} (\bibinfo {year} {2000})},\ \Eprint
  {http://arxiv.org/abs/hep-ph/9905253} {arXiv:hep-ph/9905253 [hep-ph]}
  \BibitemShut {NoStop}%
\bibitem [{\citenamefont {Ebert}\ and\ \citenamefont
  {Klimenko}(2003)}]{ebert2003quark}%
  \BibitemOpen
  \bibfield  {author} {\bibinfo {author} {\bibfnamefont {D.}~\bibnamefont
  {Ebert}}\ and\ \bibinfo {author} {\bibfnamefont {K.}~\bibnamefont
  {Klimenko}},\ }\href@noop {} {\bibfield  {journal} {\bibinfo  {journal}
  {Nuclear Physics A}\ }\textbf {\bibinfo {volume} {728}},\ \bibinfo {pages}
  {203} (\bibinfo {year} {2003})}\BibitemShut {NoStop}%
\bibitem [{\citenamefont {Ferrari}\ \emph {et~al.}(2012)\citenamefont
  {Ferrari}, \citenamefont {Garcia},\ and\ \citenamefont
  {Pinto}}]{ferrari2012chiral}%
  \BibitemOpen
  \bibfield  {author} {\bibinfo {author} {\bibfnamefont {G.~N.}\ \bibnamefont
  {Ferrari}}, \bibinfo {author} {\bibfnamefont {A.~F.}\ \bibnamefont {Garcia}},
  \ and\ \bibinfo {author} {\bibfnamefont {M.~B.}\ \bibnamefont {Pinto}},\
  }\href@noop {} {\bibfield  {journal} {\bibinfo  {journal} {Physical Review
  D}\ }\textbf {\bibinfo {volume} {86}},\ \bibinfo {pages} {096005} (\bibinfo
  {year} {2012})}\BibitemShut {NoStop}%
\bibitem [{\citenamefont {Dexheimer}\ \emph {et~al.}(2014)\citenamefont
  {Dexheimer}, \citenamefont {Menezes},\ and\ \citenamefont
  {Strickland}}]{Dexheimer:2012mk}%
  \BibitemOpen
  \bibfield  {author} {\bibinfo {author} {\bibfnamefont {V.}~\bibnamefont
  {Dexheimer}}, \bibinfo {author} {\bibfnamefont {D.~P.}\ \bibnamefont
  {Menezes}}, \ and\ \bibinfo {author} {\bibfnamefont {M.}~\bibnamefont
  {Strickland}},\ }\href {\doibase 10.1088/0954-3899/41/1/015203} {\bibfield
  {journal} {\bibinfo  {journal} {J. Phys.}\ }\textbf {\bibinfo {volume}
  {G41}},\ \bibinfo {pages} {015203} (\bibinfo {year} {2014})},\ \Eprint
  {http://arxiv.org/abs/1210.4526} {arXiv:1210.4526 [nucl-th]} \BibitemShut
  {NoStop}%
\bibitem [{\citenamefont {Pili}\ \emph {et~al.}(2014)\citenamefont {Pili},
  \citenamefont {Bucciantini},\ and\ \citenamefont {Del~Zanna}}]{Pili:2014npa}%
  \BibitemOpen
  \bibfield  {author} {\bibinfo {author} {\bibfnamefont {A.~G.}\ \bibnamefont
  {Pili}}, \bibinfo {author} {\bibfnamefont {N.}~\bibnamefont {Bucciantini}}, \
  and\ \bibinfo {author} {\bibfnamefont {L.}~\bibnamefont {Del~Zanna}},\ }\href
  {\doibase 10.1093/mnras/stu215} {\bibfield  {journal} {\bibinfo  {journal}
  {Mon. Not. Roy. Astron. Soc.}\ }\textbf {\bibinfo {volume} {439}},\ \bibinfo
  {pages} {3541} (\bibinfo {year} {2014})},\ \Eprint
  {http://arxiv.org/abs/1401.4308} {arXiv:1401.4308 [astro-ph.HE]} \BibitemShut
  {NoStop}%
\bibitem [{\citenamefont {Frieben}\ and\ \citenamefont
  {Rezzolla}(2012)}]{Frieben:2012dz}%
  \BibitemOpen
  \bibfield  {author} {\bibinfo {author} {\bibfnamefont {J.}~\bibnamefont
  {Frieben}}\ and\ \bibinfo {author} {\bibfnamefont {L.}~\bibnamefont
  {Rezzolla}},\ }\href {\doibase 10.1111/j.1365-2966.2012.22027.x} {\bibfield
  {journal} {\bibinfo  {journal} {Mon. Not. Roy. Astron. Soc.}\ }\textbf
  {\bibinfo {volume} {427}},\ \bibinfo {pages} {3406} (\bibinfo {year}
  {2012})},\ \Eprint {http://arxiv.org/abs/1207.4035} {arXiv:1207.4035 [gr-qc]}
  \BibitemShut {NoStop}%
\bibitem [{\citenamefont {Mastrano}\ \emph {et~al.}(2015)\citenamefont
  {Mastrano}, \citenamefont {Suvorov},\ and\ \citenamefont
  {Melatos}}]{Mastrano:2015rfa}%
  \BibitemOpen
  \bibfield  {author} {\bibinfo {author} {\bibfnamefont {A.}~\bibnamefont
  {Mastrano}}, \bibinfo {author} {\bibfnamefont {A.~G.}\ \bibnamefont
  {Suvorov}}, \ and\ \bibinfo {author} {\bibfnamefont {A.}~\bibnamefont
  {Melatos}},\ }\href {\doibase 10.1093/mnras/stu2671} {\bibfield  {journal}
  {\bibinfo  {journal} {Mon. Not. Roy. Astron. Soc.}\ }\textbf {\bibinfo
  {volume} {447}},\ \bibinfo {pages} {3475} (\bibinfo {year} {2015})},\ \Eprint
  {http://arxiv.org/abs/1501.01134} {arXiv:1501.01134 [astro-ph.HE]}
  \BibitemShut {NoStop}%
\end{thebibliography}%

\end{document}